# Analysis of Memory Ballooning Technique for Dynamic Memory Management of Virtual Machines (VMs)

A B M Moniruzzaman
*Department of Computer Science and Engineering,*
*Daffodil International University*
*abm.mzkhan@gmail.com*

*Abstract*

*Memory ballooning is dynamic memory management technique for virtual machines (VMs). Ballooning is a part of memory reclamation technique operations used by a hypervisor to allow the physical host system to retrieve unused memory from certain guest virtual machines (VMs) and share it with others. Memory ballooning allows the total amount of RAM required by guest VMs to exceed the amount of physical RAM available on the host. Memory overcommitment enables a higher consolidation ratio in a hypervisor. Using memory overcommitment, users can consolidate VMs on a physical machine such that physical resources are utilized in an optimal manner while delivering good performance. Hence memory reclamation is an integral component of memory overcommitment. In this paper, we address that the basic cause of memory that ballooning is memory overcommitment from using memory-intensive virtual machines. We compared to others reclamation technique and identify Cost Associate with Memory Ballooning in state of Memory Overcommitment. The objective of this paper is to analyse memory ballooning technique for dynamic memory management of VMs. For this analysis, VMware based virtualization software e.g ESXi Server, vCenter Server, vSphere Client are installed and configured on the Centre for Innovation and Technology (CIT) Lab, DIU; for monitor and analyze VM performance for memory ballooning technique. The performance of memory ballooning technique is evaluated with two different test cases. The purpose is to help users understand, how this technique impact the performance. Finally, we presents the throuhput of heavy workload with different memory limits when using ballooning or swapping; and analyse VM performance issue for this technique.*

***Keywords:*** *Memory ballooning, memory reclamation technique, Virtual machines (VMs), memory overcommitment, hypervisor, virtual memory.*

## 1. Introduction

Virtualization promises to increase efficiency by enabling workload consolidation [5], [6], [7], [8]. It enables users to consolidate virtual hardware on less physical hardware, thereby efficiently using hardware resources. The consolidation ratio is a measure of the virtual hardware that has been placed on physical hardware [14]. A higher consolidation ratio typically indicates greater efficiency. Memory overcommitment raises the consolidation ratio [4], increases operational efficiency, and lowers total cost of operating virtual machines. A virtual machine (VM) requires resources from a hypervisor [9], one of which is memory. When multiple VMs are on a physical server and require more resources than are available, a shortage occurs and VMs compete for resources [15]. In the case of a memory shortage, hypervisor has a number of mechanisms in place to handle memory shortages and limit the amount of performance degradation the VM experiences. These include include transparent page sharing, memory ballooning, and as a last resort, swapping guest memory to disk.

Virtual memory ballooning is a computer memory reclamation technique used by a hypervisor to allow the physical host system to retrieve unused memory from certain guest virtual machines (VMs) and share it with others. Memory ballooning allows the total amount of RAM required by guest VMs to exceed the amount of physical RAM available on the host. When the host system runs low on physical RAM resources, memory ballooning allocates it selectively to VMs.

## 2. Description

Typically, the hypervisor inflates the virtual machine balloon when it is under memory pressure [17]. By inflating the balloon, a virtual machine consumes less physical memory on the host, but more physical memory inside the guest. As a result, the hypervisor offloads some of its memory overload to the guest operating system while slightly loading the virtual machine. That is, the hypervisor transfers the memory pressure from the host to the virtual machine. Ballooning induces guest memory pressure. In response, the balloon driver allocates and pins guest physical memory. The guest operating system determines if it needs to page out guest physical memory to satisfy the balloon driver's allocation requests. If the virtual machine has plenty of free guest physical memory, inflating the balloon will induce no paging and will not impact guest performance. In this case, as illustrated in Figure 2.1, the balloon driver allocates the free guest physical memory from the guest free list. Hence, guest-level paging is not necessary.

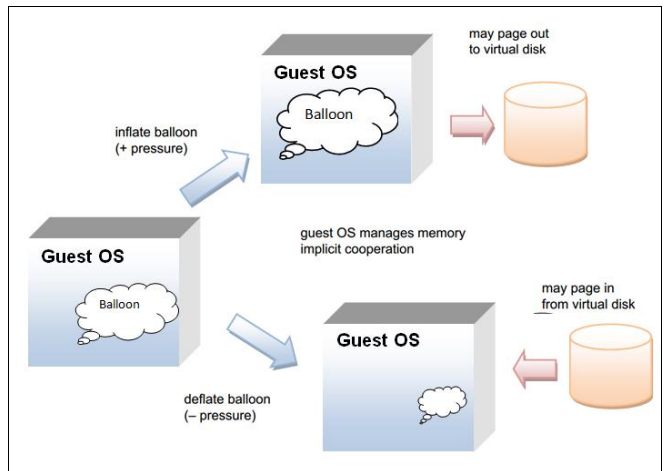

Figure 2.1: Memory Ballooning

However, if the guest is already under memory pressure, the guest operating system decides which guest physical pages to be paged out to the virtual swap device in order to satisfy the balloon driver's allocation requests. The genius of ballooning is that it allows the guest operating system to intelligently make the hard decision about which pages to be paged out without the hypervisor's involvement.

If a VM only uses a portion of the memory that it was allocated, the ballooning technique makes it available for the host to use. For example, if all the VMs on a host are allocated 8 GB of memory, some of the VMs will only use half the allotted share. Meanwhile, one VM might need 12 GB of memory for an intensive process. Memory ballooning allows the host to borrow that unused memory and allocate it to the VMs with higher memory demand. The guest operating system runs inside the VM, which is allocated a portion of memory. Therefore, the guest OS is unaware of the total memory available. Memory ballooning makes the guest operating system aware of the host's memory shortage.

Due to the virtual machine's isolation, the guest operating system is not aware that it is running inside a virtual machine and is not aware of the states of other virtual machines on the same host [13]. When the hypervisor runs multiple virtual machines and the total amount of the free host memory becomes low, none of the virtual machines will free guest physical memory because the guest operating system cannot detect the host's memory shortage. Ballooning makes the guest operating system aware of the low memory status of the host.

Ballooning is a part of normal operations when memory is overcommitted. The fact that ballooning occurrence is not necessarily an indication of a performance problem. The use of the balloon driver enables the guest to give up physical memory pages that are not being used. In fact, ballooning can be a sign that you're getting extra value out of the memory you have in the host. However, if ballooning causes the guest to give up memory that it actually needs, performance problems can occur due to guest operating system paging. Note, however, that this is fairly uncommon because the guest operating system will always assign already-free memory to the balloon driver whenever possible, thereby avoiding any guest operating system swapping.

### 3.2. Compared to others reclamation technique

Virtualization providers such as hypervisor enable memory ballooning. VMware memory ballooning, Microsoft Hyper-V dynamic memory, and the open source KVM balloon process are similar in concept. The host uses balloon drivers running on the VMs to determine how much memory it can take back from an under-utilizing VM [18]. Balloon drivers must be installed on any VM that participates in the memory ballooning technique. Contemporary hypervisors such as VMware ESX [1], Hyper-V [10], KVM [11], and Xen [12] implement diffrent memory overcommitment, reclamation, and optimization strategies.

| Technique | ESX | Hyper-V | KVM | Xen |
|---|---|---|---|---|
| Share | Yes | - | Yes | - |
| Balloon | Yes | Yes | Yes | Yes |
| Compress | Yes | - | - | - |
| Hypervisor swap | Yes | Yes | Yes | - |
| Memory hot-add | - | Yes | - | - |
| Transcendent Memory | - | - | - | Yes |

Table 1: Comparing memory overcommitment technologies in existing hypervisors

Table 1 summarizes the memory reclamation technologies implemented in existing hypervisors; also comparing memory overcommitment technologies in existing hypervisors. Memory Ballooning technique is used by exiting popular hypervisors.

### 3.3. Cost Associate with Memory Ballooning in state of Memory Overcommitment

Cost of Memory Overcommitment incurs certain cost in terms of compute resource as well as VM performance [16]. This section provides a qualitative understanding of the diffrent sources of cost and their magnitude.

| Cost | Yes/No | Cause/when happen |
| --- | --- | --- |
| Reclamation cost | Yes | balloon driver expands |
| CPU cost | Yes | memory allocation and reclamation inside the VM |
| Storage cost | Yes | guest OS may swap out memory pages to the guest swap space. This incurs storage space and storage bandwidth cost. |
| Page-fault cost (CPU cost) | Yes | A ballooned page acquired by the balloon driver may subsequently be released by it. The guest OS or application may then allocate and access it. This incurs a page-fault in the guest OS as well as ESX. The page-fault incurs a low CPU cost since a memory page simply needs to be allocated. |
| Page-fault cost (Storage Cost) | Yes | During reclamation by ballooning, application pages may have been swapped out by the guest OS. When the application attempts to access that page, the guest OS needs to swap it in. This incurs a storage bandwidth cost. |
| Wait cost | Yes | A temporal wait cost may be incurred by application if its pages were swapped out by the guest OS. The wait cost of swapping in a memory page by the guest OS incurs a smaller overall wait cost to the application than a hypervisor-level swap-in. This is because during a page fault in the guest OS, by one thread, the guest OS may schedule another thread. However, if ESX is swapping in a page, then it may deschedule the entire VM. This is because ESX cannot reschedule guest OS threads. |

Table 02: Cost Associate with Memory Ballooning in state of Memory Overcommitment

When hypervisor is memory overcommitted and powered-on VMs attempt to consume more memory than hypervisor memory, then hypervisor will begin to actively reclaim memory from VMs. Hence memory reclamation is an integral component of memory overcommitment. Table 02 shows the ballooning technique and its associated cost.

## 4. Experimental Environment Setup

A wide range of visualization solutions have been deployed. ESX, Hyper-V, KVM and Xen are the most popular ones [17]. VMware ESX Server is a thin software layer designed to multiplex hardware resources efficiently among virtual machines running unmodified commodity operating systems [1]. The VMware vSphere™ Client exposes several memory statistics in the performance charts. Among them are charts for the following memory types: consumed, active, shared, granted, overhead, balloon, swapped, and compressed [2].

We install and run ESXi Server, a 4-6 of heavy loaded Virtual Machines, a vCenter Server, and a working installation of vSphere Client. We tested on a HP ProLiant DL360 G7 server with 32GB of physical memory and vSphere 4.1. Balloon activity can be monitored in Performance Chart through vSphere Client. The metric here follow in this case is Balloon Average in kilobytes. We select the same metric when monitor the Ballooning activity for ESXi as well.

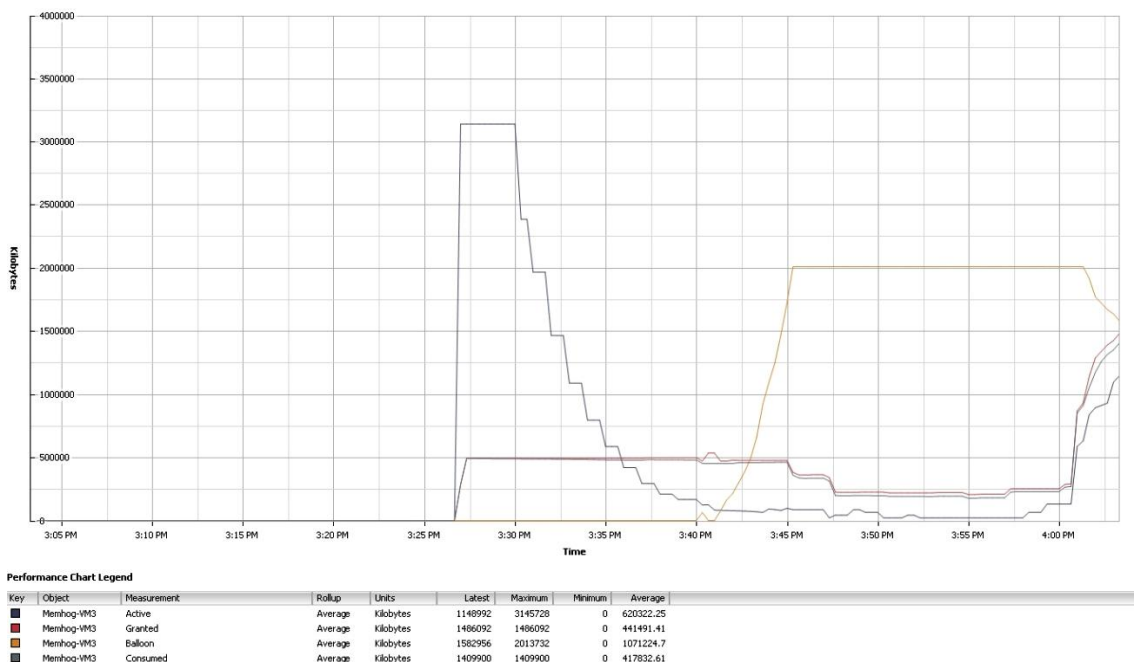

Figure: 3.1 Memory Ballooning on ESXi.

To monitor the Ballooning activity using vSphere client for individual VM, follow these steps: Open vSphere Client  then Log in to the vCenter Server after that go to the Home Screen, select VMs and Templates. Choose the VM that is required heavy memory loaded, and go to the Performance tab, and then switch to the Memory screen. Then click on the Advanced tab, and select Chart Options; Select Balloon as the metric and click on OK. In the figure 3.1 there is Ballooning activity on VM.

To monitor the Ballooning activity using vSphere client for your ESXi, you should follow these steps: Open vSphere Client then Log in to the vCenter Server. On your Home Screen, select Hosts and Clusters. Choose the poorly performing ESXi host. After that, go to the Performance tab and switch to the Memory screen. Click on the Advanced tab and select Chart Options. Select Balloon as the metric.Click on OK. Figure 3.2 shows ballooning activity on your ESXi host. In this figure 3.2, this ESXi host is involved in ballooning and its VMs are actively releasing inactive memory pages.

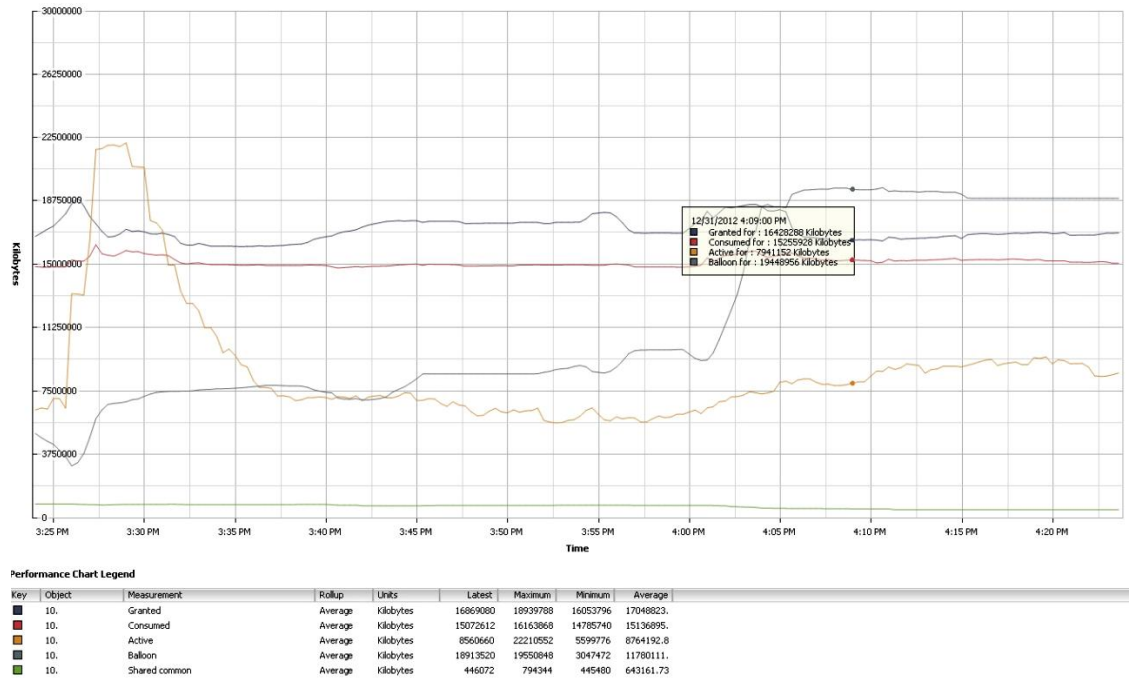

Figure 3.2: Ballooning activity on ESXi host

In the vSphere Client, use the Memory Balloon metric to monitor a host's ballooning activity. This metric represents the total amount of memory claimed by the balloon drivers of the virtual machines on the host. The memory claimed by the balloon drivers can be used by other virtual machines. Again, this is not a performance problem, but it represents that the host starts to take memory from less needful virtual machines for those with large amounts of active memory. If the host is ballooning, check the swap rate counters (Memory Swap In Rate and Memory Swap Out Rate), which might indicate performance problems, but it does not mean that you have a performance problem presently. It means that the unallocated pRAM on the host has dropped below a predefined threshold.

## 5. Test Cases for Performance Evaluation of Memory Ballooning

In this section, the performance of memory ballooning technique is evaluated with two different test cases. The purpose is to help users understand how this technique, impact the performance of dynamic virtual memory management for VMs. First test case, we ran the test with Memory Ballooning enabled to see how much memory was reclaimed through this technique along with swapping. Second test case, we tested the throughput of the kernel compile workload with different memory limits when using ballooning or swapping. This experiment was contrived to use only ballooning or swapping technique with evaluation chart.

First test case: We started a server load with about 10GB of free memory remaining after loading VMs. After that, we created extra demand for 9GB plus of memory into the VMs in order to create memory shortage. This causes force ESX to reclaim memory. Memory pressure from this action theoretically requires no more than 9.5GB of Free ESX Memory and reclaimed memory combined. However, our test results show that the hypervisor reclaimed about 9GB of memory while additionally losing 8GB of Free Memory. This appears to imply that ESX requires 17GB of memory for the 9.5 GB injection as shown in the following chart:

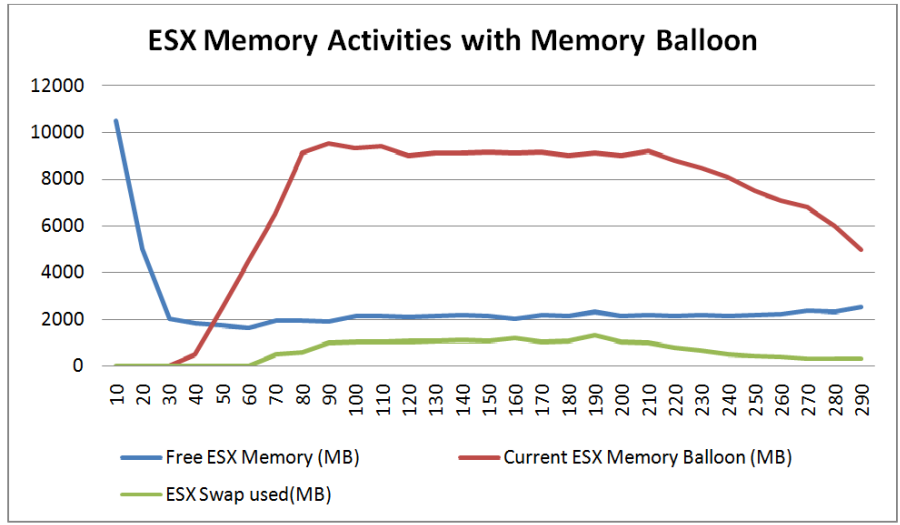

Figure 5.1: ESX Memory Activities with Memory Balloon

Therefore, we ran the same test with Memory Ballooning disabled to see how much memory was reclaimed through kernel swapping only. The result shows the exact amount of memory that ESX needs to reclaim. Our test result, in figure 2, shows 8GB of Free ESX memory consumed as well as 1.9GB of memory reclaimed through kernel swapping. These figures, adding to roughly 10 GB, are much closer to the values we originally expected.

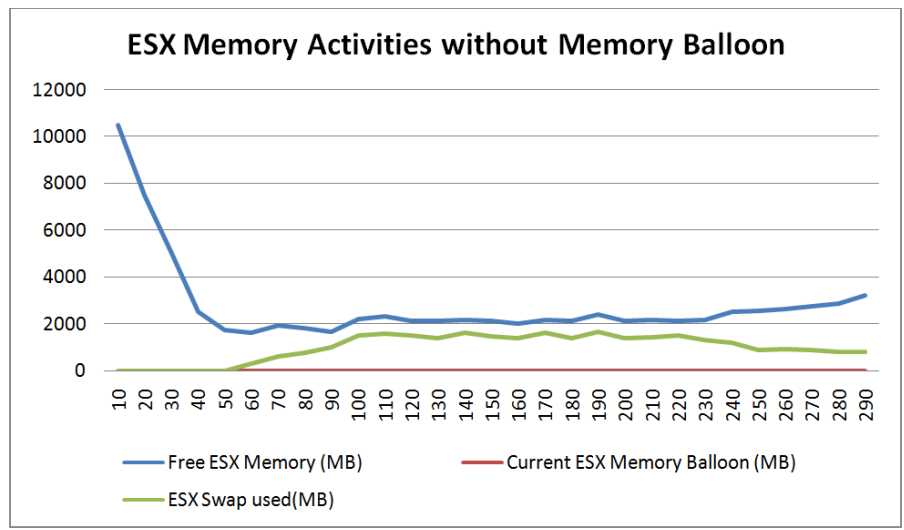

Figure 5.2: ESX Memory Activities without Memory Ballooning

Analysis from test case 01: Based on the test case 01 findings, it appeared that the current Ballooned Memory only indicated the amount of the guest memory that was pinned but not necessarily the amount memory physically reclaimed by the ESX. VMware installs a balloon driver inside the guest OS and signals the driver to begin to "inflate" when it begins to encounter contention for machine memory, defined as the amount of free machine memory available for new guest machine allocation requests dropping low level. Changes in memory consumed are inversely proportional to changes in memory reclaimed [3] and only the "Consumed" memory is backed by physical memory [2]. In order to maximize the ability of ESXi to recover idle memory from virtual machines, the balloon driver should be enabled on all virtual machines. The balloon driver should never be deliberately disabled on a virtual machine. Disabling the balloon driver might cause unintended performance problems.

Test Case 02: In this test case experiment, we forced to change virtual machine's memory limit value from the default in the VM memory reclamation state. We turned off Page Sharing in order to isolate the performance impact of ballooning or swapping. When, the host memory is much larger than the virtual machine memory size, the host free memory goes into the high state. In this time, automatically ESX uses ballooning to reclaim memory.

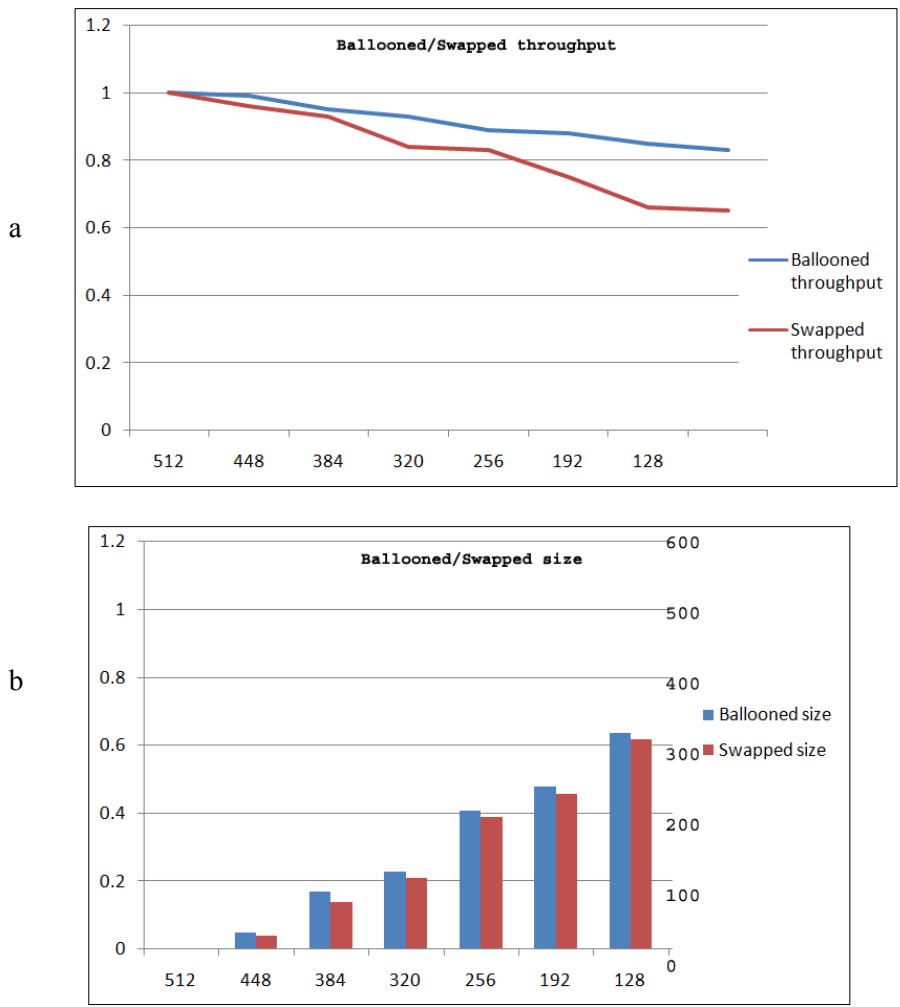

Figure 5.3 (a, b): Throughput of the kernel compile workload with different memory limits when using ballooning or swapping

After that observation and data collection, ballooning is turned off to observe the performance of using swapping only. The ballooned and swapped memory sizes were also collected when the virtual machine ran steadily.

The figure 5.3, presents the throughput of the kernel compile workload with different memory limits when using ballooning or swapping. This experiment was contrived to use only ballooning or swapping, not both. While this case will not often occur in production environments, it shows the performance penalty due to either technology on its own. The throughput is normalized to the case where virtual machine memory is not reclaimed

Analysis from test case 02: The kernel compile workload has very little memory reuse and most of the guest physical memory is used as buffer caches for the kernel source files. With ballooning, the guest operating system reclaims guest physical memory upon the balloon driver's allocation request by dropping the buffer pages instead of paging them out to the guest virtual swap device. Because dropped buffer pages are not reused frequently, the performance impact of using ballooning is trivial. However, with hypervisor swapping, the selected guest buffer pages are unnecessarily swapped out to the host swap device and some guest kernel pages are swapped out occasionally, making the performance of the virtual machine degrade when the memory limit decreases. When the memory limit is very low, the throughput loss is about 34 percent in the swapping case. Balloon inflation is a better approach to memory reclamation from a performance perspective.

High-usage values usually do not cause performance degradation. A consistently high memory usage value (94% or greater) indicates that the host is probably lacking the memory required to meet the demand. If the active memory size is the same as the granted memory size, the demand for memory is greater than the memory resources available. If the active memory is consistently low, the memory size might be too large. If the memory usage value is high, and the host has high ballooning or swapping, check the amount of free physical memory on the host. A free memory value of 6% or less indicates that the host cannot handle the demand for memory. This leads to memory reclamation, which might degrade performance. If the host has enough free memory, check the resource shares, reservation, and limit settings of the virtual machines and resource pools on the host. Verify that the host settings are adequate and not lower than those set for the virtual machines.

If the host has little free memory available, or if you notice a degradation in performance, consider taking the following actions. Reduce the memory space on the virtual machine, and correct the cache size if it is too large. This frees up memory for other virtual machines. If the memory reservation of the virtual machine is set to a value much higher than its active memory, decrease the reservation setting so that the VMkernel can reclaim the idle memory for other virtual machines on the host. Migrate one or more virtual machines to a host in a DRS cluster. Add physical memory to the host.

## 6. Conclusion

In this paper, we compared to others reclamation technique and identify Cost Associate with Memory Ballooning in state of Memory Overcommitment. We evaluate targeting to analyse memory ballooning technique for dynamic memory management of VMs. For this analysis, VMware based virtualization software e.g ESXi Server, vCenter Server, vSphere Client are installed and configured on the Centre for Innovation and Technology (CIT) Lab, DIU; for monitor and analyze VM performance for memory ballooning technique. The performance of memory ballooning technique is evaluated with two different test cases. The purpose is to help users understand, how this technique impact the performance. Finally, we presents the throughput of heavy workload with different memory limits when using ballooning or swapping; and analyse VM performance issue for this technique.

Analysis from test case 01, the balloon driver should never be deliberately disabled on a virtual machine. Disabling the balloon driver might cause unintended performance problems. Analysis from test case 02, when, the host memory is much larger than the virtual machine memory size, the host free memory goes into the high state. In this time, automatically ESX uses ballooning to reclaim memory. With ballooning, the guest operating system reclaims guest physical memory upon the balloon driver's allocation request by dropping the buffer pages instead of paging them out to the guest virtual swap device. Because dropped buffer pages are not reused frequently, the performance impact of using ballooning is trivial.

To ensure best performance, the host memory must be large enough to accommodate the active memory of the virtual machines. The active memory can be smaller than the virtual machine memory size. This allows you to over-provision memory, but still ensures that the virtual machine active memory is smaller than the host memory. The basic cause of memory that ballooning is again memory overcommitment from using memory-intensive virtual machines. However, this is just indicative, which means that the presence of ballooning does not always say it's a performance problem.

# Author

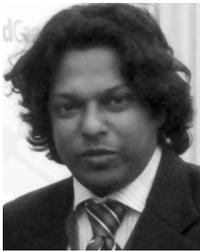

**A B M Moniruzzaman** Received his B.Sc (Hon's) degree in Computing and Information System (CIS) from London Metropolitan University, London, UK and M.Sc degree in Computer Science and Engineering (CSE) from Daffodil International University, Dhaka, Bangladesh in 2005 and 2013, respectively. Currently he is working as a Lecturer of the department of Computer Science and Engineering ad Daffodil International University. He is also working on research on Cloud Computing and Big Data Analytics as a research associate at RCST (Research Center for Science and Technology) at Daffodil International University (DIU), Dhaka, Bangladesh. Besides, his voluntarily works as reviewer of many international journals including IEEE, Elsevier, IGI-Global. He has got 7(seven) international publications including journals and proceedings. He is a student member of IEEE. His research interests include Cloud Computing, Cloud Applications, Open-source Clouds, Cloud Management Platforms, Building Private and Hybrid Cloud with FOSS software, Big Data Management, Agile Software Development, Hadoop, MapReduce, Parallel and Distributed Computing, Clustering, High performance computing, Distributed Databases, NoSQL Databases.